\documentclass[9pt,twocolumn,twoside]{osajnl}

\journal{ol}

\setboolean{shortarticle}{true} 
\usepackage[utf8]{inputenc}
\usepackage[T1]{fontenc}
\usepackage{textcomp}

\title{Mid-infrared frequency comb generation via cascaded quadratic nonlinearities in quasi-phase-matched waveguides}

\author[1,*]{Abijith S. Kowligy}
\author[1,2]{Alex Lind}
\author[1]{Daniel D. Hickstein}
\author[1]{David R. Carlson}
\author[1]{Henry Timmers}
\author[3]{Nima Nader}
\author[1,4]{Flavio C. Cruz}
\author[3]{Gabriel Ycas}
\author[1]{Scott B. Papp}
\author[1,2]{Scott A. Diddams}

\affil[1]{Time and Frequency Division, National Institute of Standards and Technology, 325 Broadway, Boulder CO, 80305}
\affil[2]{Department of Physics, University of Colorado, Boulder CO, 80305}
\affil[3]{Applied Physics Division, National Institute of Standards and Technology, 325 Broadway, Boulder CO, 80305}
\affil[4]{Instituto de Fisica Gleb Wataghin, Universidade Estadual de Campinas, Campinas, SP, 13083-859, Brazil}
\affil[*]{Corresponding author: abijith.kowligy@gmail.com}

\ociscodes{(190.0190) Nonlinear optics; (130.3060) Infrared, Integrated optics; (320.6629) Supercontinuum generation, Ultrafast optics.}

\begin{abstract}
We experimentally demonstrate a simple configuration for mid-infrared (MIR) frequency comb generation in quasi-phase-matched lithium niobate waveguides using the cascaded-$\chi^{(2)}$ nonlinearity. With nanojoule-scale pulses from an Er:fiber laser, we observe octave-spanning supercontinuum in the near-infrared with dispersive-wave generation in the 2.5--3 \textmu{}m region and intra-pulse difference-frequency generation in the 4--5 \textmu{}m region. By engineering the quasi-phase-matched grating profiles, tunable, narrow-band MIR and broadband MIR spectra are both observed in this geometry. Finally, we perform numerical modeling using a nonlinear envelope equation, which shows good quantitative agreement with the experiment---and can be used to inform waveguide designs to tailor the MIR frequency combs. Our results identify a path to a simple single-branch approach to mid-infrared frequency comb generation in a compact platform using commercial Er:fiber technology.
\end{abstract}

\setboolean{displaycopyright}{true}

\begin{document}

\maketitle

\begin{figure}[th!]
\centering
\includegraphics[width=0.875\linewidth]{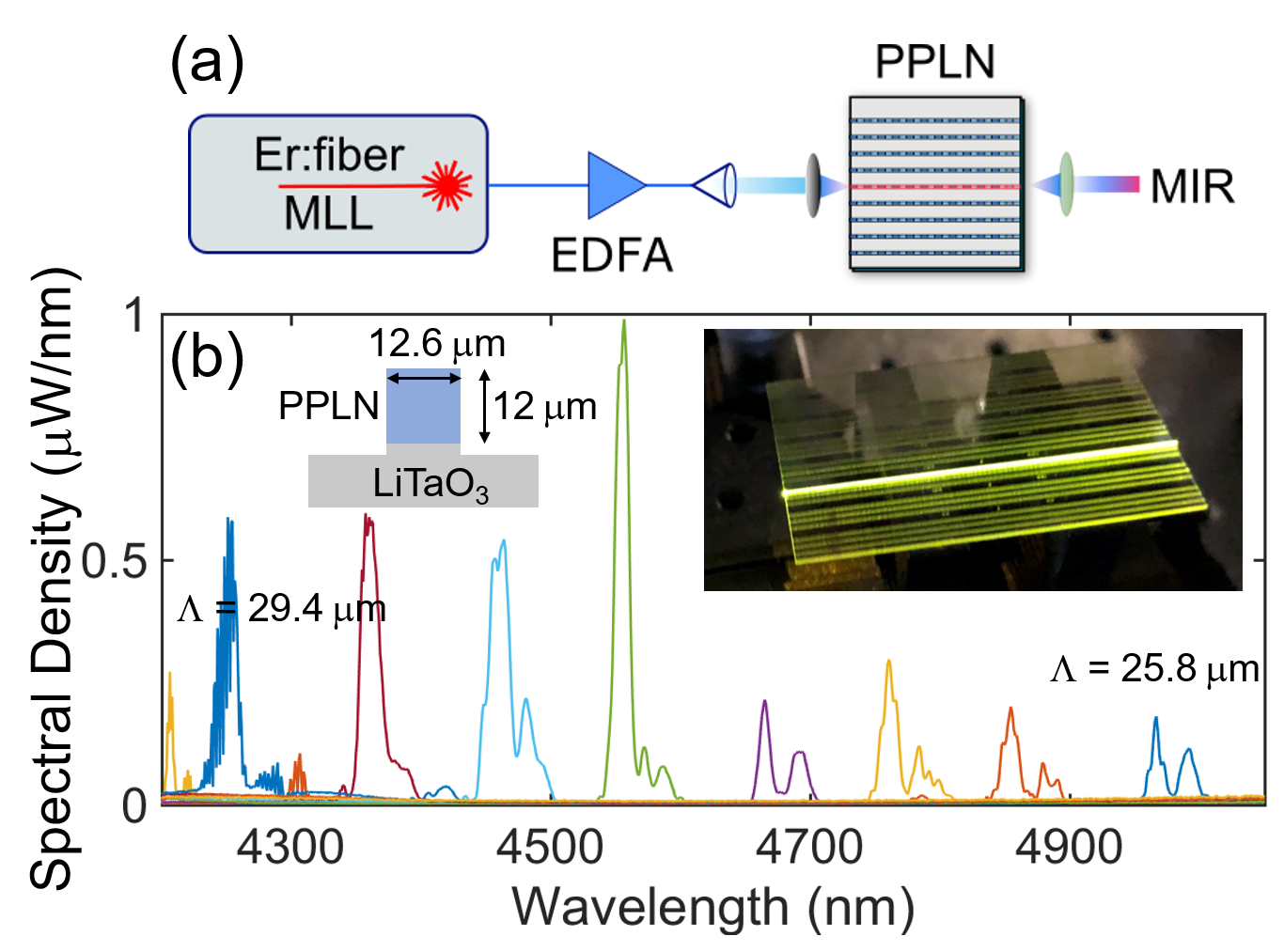}
\caption[Experimental~Setup]{(a) The femtosecond pulses from a 1.5~\textmu m Er:fiber mode-locked laser (MLL) are amplified with an erbium-doped fiber amplifier (EDFA), and focused into periodically poled lithium niobate (PPLN) waveguides. MIR light that is generated in the waveguide is sent to a spectrometer or photodetector. (b)  DFG spectra from the waveguide-chip as the poling periods are tuned from $\Lambda = 25.8$~\textmu{}m to $\Lambda=29.4$~\textmu m. (Inset left): The small cross section of the PPLN waveguide provides strong optical confinement and high intensity. (Inset right): The waveguide glows with visible light as a result of parasitic sum-frequency generation.
\label{fig:setup}
}
\end{figure}

Mid-infrared (3--25 \textmu m) frequency combs are desirable for many multidisciplinary scientific goals including precision spectroscopy in the molecular fingerprint region \cite{schliesser_mid-infrared_2012}, referencing quantum cascade lasers (QCL) \cite{knabe_absolute_2013}, probing fundamental symmetries in physics \cite{stoeffler_high_2011}, and novel imaging techniques \cite{huber_ultrafast_2016}. For certain applications such as dual-comb spectroscopy \cite{coddington_dual-comb_2016} and absolute frequency metrology \cite{malara_absolute_2008}, compact and chip-scale geometries are also desirable. In the near-infrared (NIR), frequency combs have seen extensive research and development due to the robust and commercially available erbium-, ytterbium-, and thulium-doped gain fiber, whereas the MIR has been less explored \cite{Hu15,Antipov16}. Nascent technologies such as MIR QCL frequency combs have also been demonstrated \cite{Hugi2012}. In contrast, frequency conversion to the MIR using robust, stable NIR frequency combs in quadratic ($\chi^{(2)}$) and cubic ($\chi^{(3)}$) media has been appealing due to the availability of high-power amplifiers in the NIR region and widely-transparent nonlinear optical materials. Such nonlinear  techniques include parametric oscillation in $\chi^{(2)}$ and $\chi^{(3)}$ optical cavities \cite{vainio_mid-infrared_2016,wang_mid-infrared_2013,yu_modelocked_2016}, difference-frequency generation (DFG) \cite{erny_mid-infrared_2007,cruz_mid-infrared_2015}, and supercontinuum generation (SCG) \cite{hickstein_ultrabroadband_2017,herkommer_mid-infrared_2017}.

DFG, in particular, has been the workhorse of many experiments utilizing MIR frequency combs. Owing to the inherent offset-frequency subtraction in the DFG process, comb stabilization is simplified and requires only repetition rate stability \cite{Kobayashi2002}. However, conventional DFG experiments are difficult to miniaturize due to the requirement that the pump and the signal pulses must be overlapped in both space and time for high-efficiency conversion \cite{maser_coherent_2017,mayer_offset-free_2016}, which typically requires alignment optics and mechanical delay stages. In contrast, SCG requires only a single pulse, but the conversion efficiency to the MIR is limited \cite{hickstein_ultrabroadband_2017}. 

In this Letter, we experimentally demonstrate a simplified configuration for MIR frequency comb generation by combining spectral broadening and difference-frequency generation in the same nonlinear optical waveguide. In particular, we utilize the nonlinear broadening due to the cascaded-$\chi^{(2)}$ \cite{stegeman_2_1996,DeSalvo1992}  
 nonlinearity in a quasi-phase-matched (QPM) waveguide and intra-pulse difference-frequency mixing to generate MIR frequency combs. In the cascaded-$\chi^{(2)}$ process, the pump pulse undergoes strong intensity dependent phase modulation induced by phase-mismatched second-harmonic generation (SHG) \cite{DeSalvo1992}. This results in an effective, self-defocusing cubic nonlinearity and leads to spectral broadening in the normal dispersion regime. Since the phase-mismatch is controlled by the QPM grating profile, engineered quadratic and effective cubic nonlinear interactions are possible, finding both quantum and classical applications in squeezed light generation \cite{Youn96}, all-optical switching \cite{Kanter2001}, femtosecond pulse generation \cite{wise2002}, broadband SCG \cite{zhou_parametrically_2017,wang_cross-polarized_2017}, mode-locking GHz-rate solid-state lasers \cite{mayer_watt-level_2017}, and frequency comb stabilization \cite{langrock_generation_2007,phillips_supercontinuum_2011}. In previous comb-stabilization experiments, octave spanning supercontinua in the NIR were observed in reverse-proton-exchanged PPLN waveguides using Yb: and Tm:fiber lasers \cite{langrock_generation_2007,phillips_supercontinuum_2011}, but relatively high pulse energies (>10~nJ) were required. In a previous Er:fiber pumped experiment, the infrared wavelengths did not extend beyond 3~\textmu{}m \cite{langrock_generation_2007}.  

Using an Er:fiber laser pump, we demonstrate frequency combs in the 4--5 \textmu m region in two configurations: (i) tunable, offset-free, narrow-band mid-infrared light in 4-cm-long periodically poled waveguides and (ii) broadband MIR in chirped (aperiodically-poled, aPP) waveguides. In the cascaded-$\chi^{(2)}$-driven SCG, we also observe the generation of dispersive waves in the 2.5--3 \textmu m region. Finally, we verify the utility of such broadband combs for applications such as DCS in a proof-of-principle multi-heterodyne experiment using two offset-free combs in the 4.8-\textmu m region. 
\begin{figure}[t!]
\centering
\includegraphics[width=\linewidth]{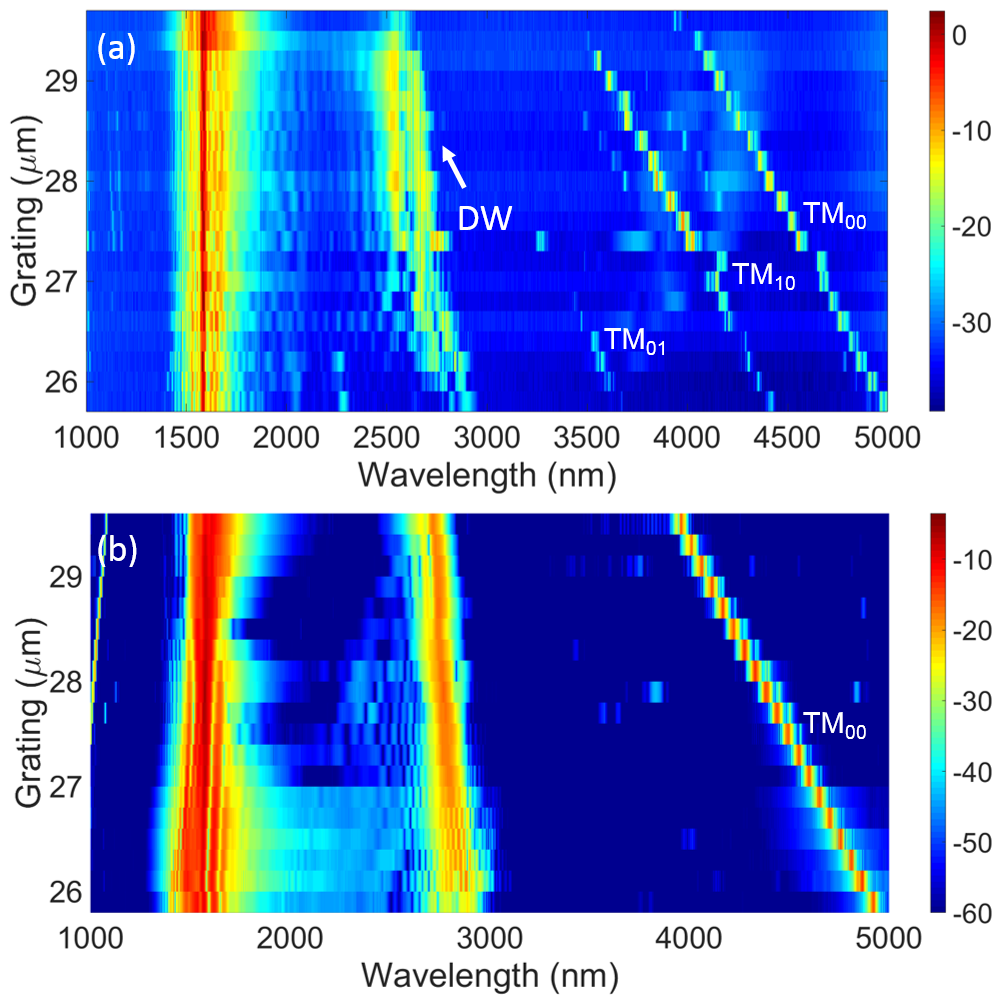}
\caption[Tunable~midIR]{(a) Experimental spectra (in logarithmic scale) for cascaded-$\chi^{(2)}$ spectral broadening and tunable, intra-pulse DFG in 4-cm-long PPLN waveguides are shown as a function of grating period. Owing to the multi-mode nature of the waveguide, the DFG also occurs in higher-order spatial modes, each of which obeys its own phase-matching condition and leads to additional discrete peaks in the mid-IR. The spectral broadening is also accompanied by dispersive-wave (DW) generation in the 2.5\textendash3 \textmu{}m wavelength region. (b) The corresponding theoretical spectra (in logarithmic scale), modeled using a single-mode nonlinear analytic envelope equation, showing good agreement with the experiment. The model only takes into account the $\mathrm{TM_{00}}$ mode, and thus the peaks corresponding to higher order spatial modes are not present. 
}
\label{fig:tunability}
\end{figure}

We use a turnkey Er:fiber laser (100 MHz repetition rate, \cite{sinclair_invited_2015}) to pump the waveguides (Fig.~\ref{fig:setup}a). The laser output is amplified with a nonlinear Er:fiber amplifier to yield 80-fs (FWHM), 2-nJ pulses. Aspheric lenses are used to couple light in and out of the 4-cm-long PPLN waveguide, exhibiting an insertion loss of 3 dB.  The waveguide-chip contains 20 waveguides with grating periods spanning 25.8\textendash29.6 \textmu m in 0.2~\textmu m increments. The waveguides are PPLN ridges on a lithium tantalate substrate \cite{nishida_direct-bonded_2003} and have cross-section dimensions of 12.6 \textmu m $\times$ 12 \textmu m (Fig.~\ref{fig:setup}b). 


With these waveguides, we observe cascaded-$\chi^{(2)}$-driven SCG in the NIR including dispersive-wave generation in the 2.5--3 \textmu{}m region due to the zero-crossing in the group-velocity dispersion (GVD, $\lambda_{ZDW} = 1.9$~\textmu{}m for bulk lithium niobate). In this process, the pump-pulse undergoes soliton fission, providing broadening in the spectral domain and temporal compression in the time-domain \cite{Dudley2006,zhou_dispersive_2015}. Simultaneously, intra-pulse DFG occurs in the waveguide, resulting in MIR light. As the grating period is changed, the MIR is smoothly tuned from 4--5 \textmu{}m (Fig.~\ref{fig:tunability}a).  Owing to the multi-mode waveguide, we also observed phase-matching to higher-order spatial modes that results in additional DFG peaks (Fig.~\ref{fig:tunability}a). Spectrally filtering and imaging the MIR on a microbolometer-array camera confirmed the presence of other spatial modes. The TM$_{00}$ DFG power is on the order of 100~\textmu W in each waveguide. The relatively long interaction length in the waveguide results in a narrow DFG phase-matching bandwidth, $\Delta\lambda_\text{FWHM} \approx 20$ nm. We note that similar DFG has been observed in bulk PPLN crystals recently and termed as DFG resonant-radiation \cite{zhou_parametrically_2017}. 

We model the nonlinear optical dynamics in the waveguide for the TM$_{00}$ mode using the single-mode nonlinear analytic envelope equation \cite{Phillips2011,conforti2010,zhou_parametrically_2017}, 
\begin{align}
\label{eq:model}
\frac{\partial A}{\partial z} + i\hat{D}A(z,t) &= i(1+\frac{i}{\omega_{0}}\frac{\partial}{\partial t}) \times \nonumber \\
& \big[\chi(z)(A^2 e^{-i\phi(z,t)} + |A|^2 e^{i\phi(z,t)}) \nonumber \\
&+\gamma (|A|^2A + A\int dt' R(t,t')|A|^2(t'))\big], 
\end{align} where $\hat{D} = \sum_{j=2} \frac{1}{j!} k_{j} (i\frac{\partial}{\partial t})^j$ is the dispersion operator, $\chi(z) = \chi^{(2)}(z)\omega_0^{2}/4\beta_{0}c^2$, $\phi(z,t) = \omega_{0} t - (\beta_{0}-\beta_{1}\omega_{0})z$, $\gamma = n_{2}\omega_{0}/c A_{\text{eff}}$ is the nonlinear Kerr parameter, and $R(t,t')$ is the Raman response function for lithium niobate \cite{bacheRaman}. For this work, we assume $d_{\text{eff}} = 19$ pm/V, $n_{2} = 2.5\times 10^{-16}$ cm$^{2}$/W,  and the Raman fraction to be $f_{R} = 0.2$ . For the $\chi^{(2)}$ nonlinearity, we take into account all the orders of the grating and use the full dispersion function, $k(\omega)$, calculated for the waveguides via COMSOL, which also yields an effective area, $A_{\text{eff}} = 75$~\textmu{}m$^{2}$, for the pump mode. By taking into account the quadratic and cubic nonlinearities of lithium niobate, the model reproduces the different spectra observed from the various waveguides (Fig.~\ref{fig:tunability}b). 

\begin{figure}[t!]
\centering
\includegraphics[width=0.95\linewidth]{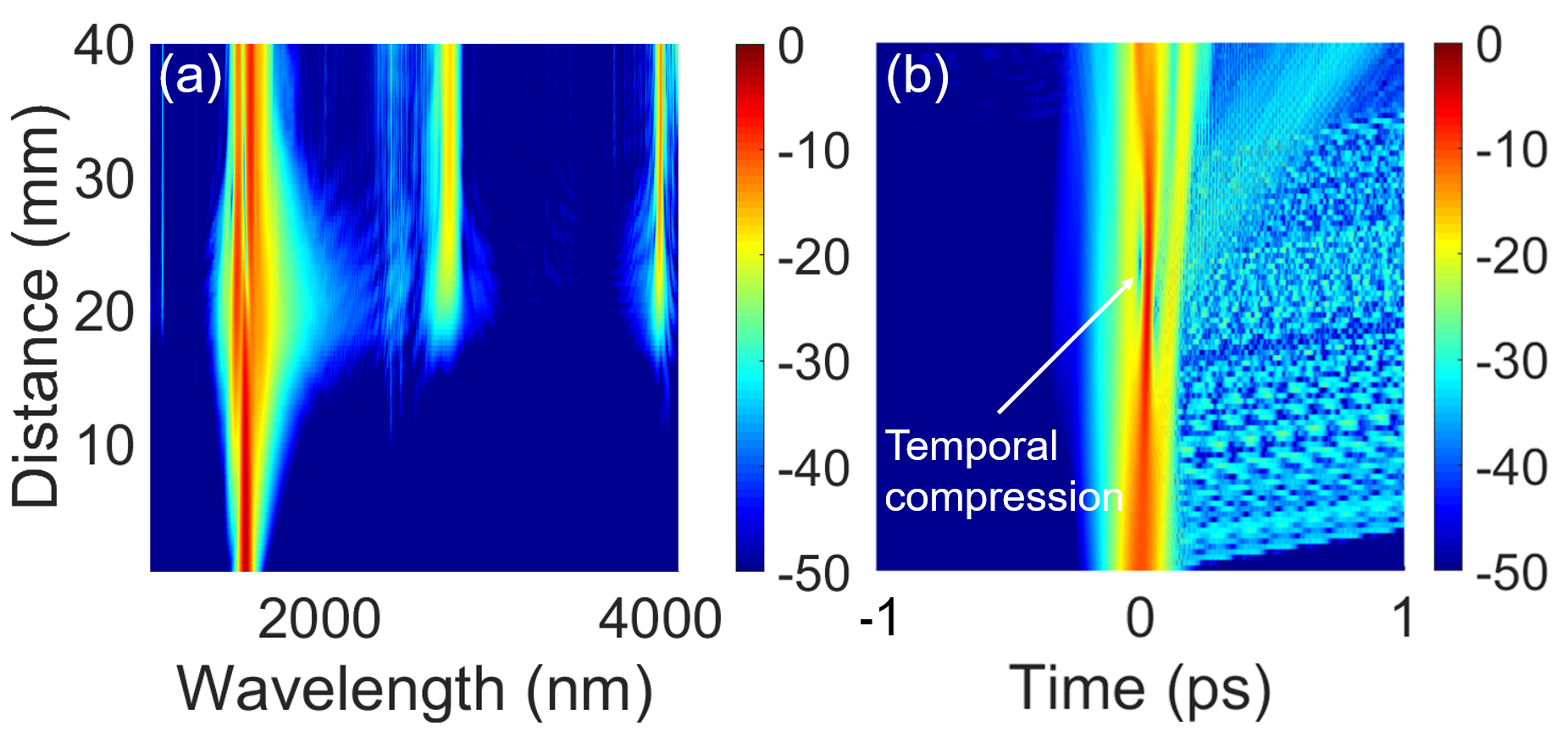}
\caption[Propagation]{(a) Spectral evolution (in logarithmic scale) as a function of distance in the 40-mm long waveguide ($\Lambda = 29.2$~\textmu{}m). The soliton fission length is approximately 15~mm. (b) The temporal evolution (in logarithmic scale) of the pulse as a function of distance in the pump frame-of-reference. Temporal compression occurs in the time-domain (minimum pulse-duration, $\tau_{FWHM} = 13$~fs). Group-velocity walkoff is observed for the DFG, limiting conversion efficiency and bandwidth.} 
\label{fig:propagation}
\end{figure}

We also study the propagation dynamics of a single waveguide ($\Lambda = 29.6$~\textmu{}m, Fig.~\ref{fig:propagation}). The observed SCG in the NIR has contributions from both the quadratic and cubic (Kerr) nonlinearities: the total cubic nonlinearity is the sum of positive n$_2$ from the Kerr nonlinearity and the negative, effective n$_2$ arising from the cascaded-$\chi^{(2)}$ effect, resulting in a net negative value. Compared to conventional SCG, where anomalous GVD ($\beta_{2} < 0$) balances the positive n$_2$ to facilitate soliton formation and fission, the observed SCG occurs in the normal dispersion regime ($\beta_{2} > 0$), balancing the negative n$_2$.  Temporal compression also occurs in the self-defocusing nonlinear dynamics \cite{bache_scaling_2007}, and corresponds to $\tau_{\text{FWHM}} \approx 13$~fs at the point of soliton fission (Fig.~\ref{fig:propagation}b). Thus, one could engineer the grating profile and waveguide dimensions to tailor the output spectra using the cascaded-$\chi^{(2)}$ nonlinearity toward sub-nanojoule-scale, few-cycle pulses \cite{Moses2006}. 

\begin{figure}[th!]
\centering
\includegraphics[width=\linewidth]{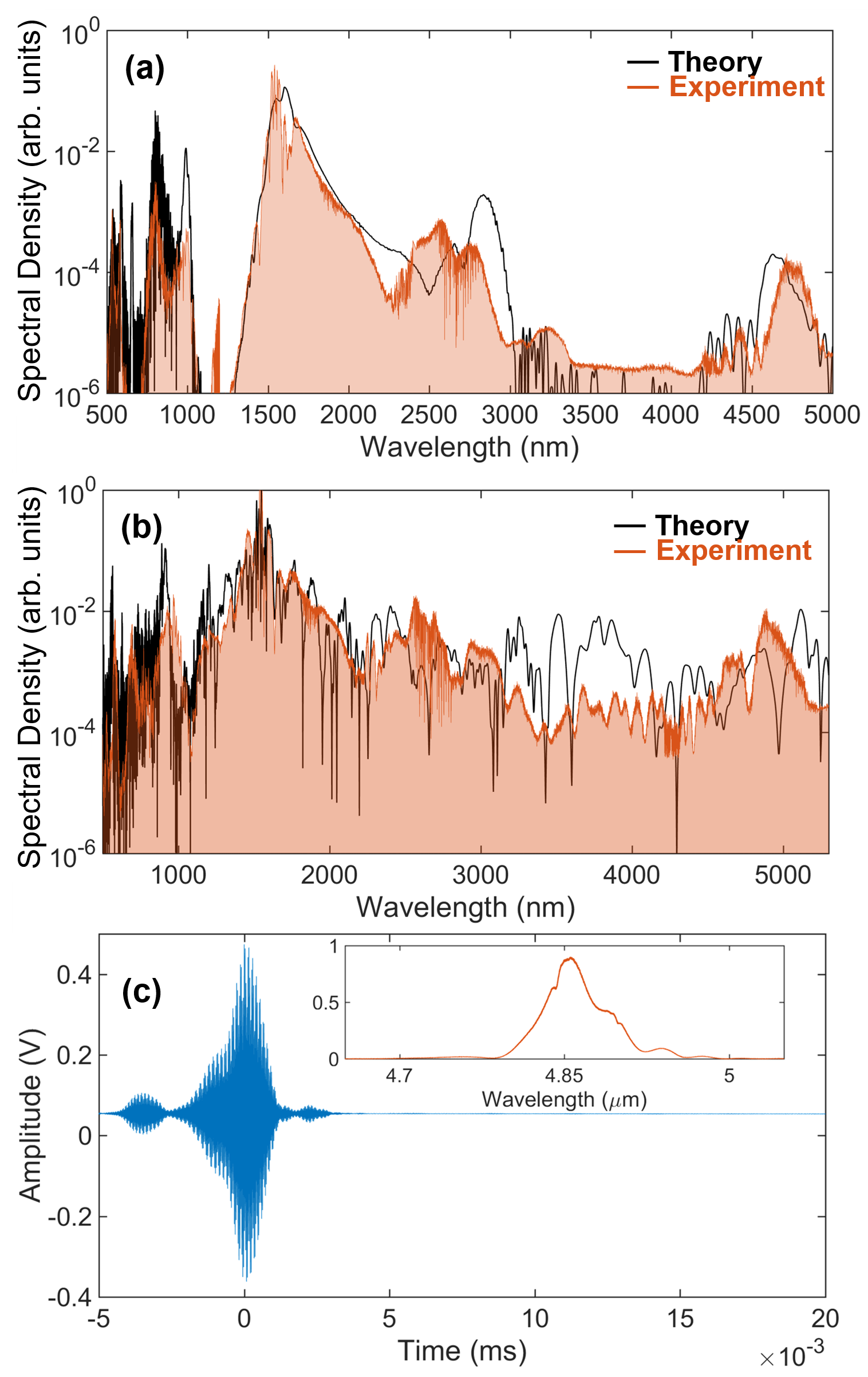}
\caption[Chirped~PPLN]{(a,b) The experimental and modeled spectra for (a) 10-mm long aPPLN waveguide, yielding broadband light in the 4.8-\textmu{}m region pumped by a 1.5-nJ, 40-fs Er:fiber pump pulse and (b)  0.5-cm long aPPLN waveguide, showing a continuum across the 0.5 -- 5 \textmu{}m decade, pumped by a 1.5-nJ, 12-fs pump pulse derived from an Er:fiber laser. (c) The center burst of the interferogram resulting from the multi-heterodyne of the two combs in (a) and (b).  (Inset): The resulting multi-heterodyne spectrum. }
\label{fig:chirpedPPLN}
\end{figure}

The ability to quasi-continuously tune the DFG across the first atmospheric window is valuable for applications such as targeted spectroscopy in molecules \cite{Nesbitt2016}. In other cases, such as dual-comb spectroscopy, broader MIR bandwidths enable spectroscopy of broadband absorbers while maintaining the frequency accuracy provided by the comb \cite{coddington_dual-comb_2016}. By employing chirped QPM grating profiles, such broadband MIR spectra can be obtained. Using aPPLN waveguides, we demonstrate the broadband DFG (Fig.~\ref{fig:chirpedPPLN}a,b), which is well predicted by the numerical modeling (Eq.~\ref{eq:model}).

We investigate two aPPLN waveguides, with cross-sectional dimensions of 15~\textmu{}m $\times$ 16~\textmu{}m, simultaneously using two Er:fiber lasers for a dual-comb (or multi-heterodyne) experiment. First, in a 10-mm-long waveguide (with a chirp in the grating from 33--29~\textmu m), the DFG light is generated with $\Delta\lambda_{\text{FWHM}} = 150$~nm around 4.8~\textmu{}m using a 40-fs, 1.5-nJ pump pulse (Fig.~\ref{fig:chirpedPPLN}a). In a second 5-mm-long waveguide (with a chirp in the grating from 29\textendash27~\textmu{}m), a decade-spanning continuum (0.5 -- 5 \textmu{}m) is generated (Fig.~\ref{fig:chirpedPPLN}b) using a few-cycle, 1.5-nJ pump pulse derived from an Er:fiber laser \cite{Timmers2017}. A chalcogenide aspheric lens and a parabolic mirror are used as output couplers for the 10-mm and 5-mm-long waveguides, respectively. 

For DCS experiments, highly coherent combs and milliwatt-scale optical powers are desirable. To demonstrate this utility of the MIR generated by the cascaded-$\chi^{(2)}$ process, we perform a proof-of-principle multi-heterodyne experiment with the spectra in Fig.~\ref{fig:chirpedPPLN}a,b. The repetition rates of the two pump lasers are locked to a microwave frequency reference and offset by $\Delta f_{\text{rep}} = 50$~Hz. The milliwatt-scale MIR spectra are spectrally filtered using a 4.5-\textmu m long-pass filter and combined on a CaF$_{2}$ beam-splitter. A liquid-nitrogen-cooled mercury-cadmium-telluride (HgCdTe) detector is used for photodetection. The time-domain signal is measured over the window, $T = 1/\Delta f_{\text{rep}} = 20$~ms, with $>350$ signal-to-noise ratio (SNR) acquired by averaging 1024 interferograms (Fig.~\ref{fig:chirpedPPLN}c). The coherent MIR waveguide output enables the high SNR and provides for 100~MHz resolution in the dual-comb spectrum (Fig.~\ref{fig:chirpedPPLN}c, inset). The DCS bandwidth is limited by the 10-mm-device MIR spectrum but can be addressed by engineering the QPM grating appropriately \cite{Phillips2010}. 

In summary, we have demonstrated a robust and straightforward technique for mid-infrared frequency comb generation in the 4--5 \textmu{}m band in quasi-phase-matched lithium niobate waveguides. Using the cascaded-$\chi^{(2)}$ nonlinearities, a single mode-locked Er:fiber laser is able to access the MIR wavelengths. By engineering the QPM grating profile, the mid-infrared spectra can be made narrowband or broadband. In addition, the DFG combs are offset-free and simplify comb-stabilization. By using two such combs, we demonstrated stable multi-heterodyne beating and presented a MIR dual-comb spectrum. We also modeled the nonlinear dynamics using a nonlinear envelope equation, which showed good agreement with the experimentally acquired spectra. This QPM-enabled engineering of quadratic and effective cubic nonlinear interactions allows for spectral engineering across multiple octaves---providing control over both the nonlinear strength and dispersion on a compact platform. Finally, higher repetition rate pump lasers can also be used for cascaded-$\chi^{(2)}$ enabled MIR generation: a 250-MHz Er:fiber laser output using only 600 pJ pump pulse energy showed similar MIR comb generation \cite{Lind17}, potentially making this approach accessible to GHz-repetition rate lasers. 


\noindent DARPA SCOUT, AFOSR (FA9550-16-1-0016), NRC, NASA, NIST. 

\noindent The authors thank Daryl Spencer and Franklyn Quinlan for helpful comments on the manuscript and Dr. Yoshiki Nishida for providing the 4-cm-long PPLN waveguide chip. This work is a contribution of the United States government and is not subject to copyright in the United States of America.

\bibliography{MIR_CascadedChi2}
\bibliographyfullrefs{MIR_CascadedChi2}
 
\end{document}